\documentclass[%
 reprint,
 amsmath,amssymb,
 aps,
floatfix,
]{revtex4-2}
\usepackage{xcolor}
\usepackage{graphicx}
\usepackage{dcolumn}
\usepackage{bm}
\usepackage[mathlines]{lineno}
\usepackage{algorithm}
\usepackage{algpseudocode}

\usepackage[english]{babel}

\usepackage{amsmath}
\usepackage{graphicx}
\usepackage{braket}
\usepackage[colorlinks=true, allcolors=blue]{hyperref}

\begin{document}

\preprint{AIP/QDBA2023}

\title{Quantum-Error-Mitigated Detectable Byzantine Agreement with Dynamical Decoupling for Distributed Quantum Computing}
\author{Matthew Prest}
 \affiliation{Columbia University - Irving Medical Center, New York, NY 10027, United States}
\author{Kuan-Cheng Chen}%
 \email{kuan-cheng.chen17@imperial.ac.uk}
\affiliation{%
 Centre for Quantum Engineering, Science and Technology (QuEST), Imperial College London, South Kensington SW7 2AZ, London,
United Kingdom 
}%

\date{\today}

\begin{abstract}
In the burgeoning domain of distributed quantum computing, achieving consensus amidst adversarial settings remains a pivotal challenge. We introduce an enhancement to the Quantum Byzantine Agreement (QBA) protocol, uniquely incorporating advanced error mitigation techniques: Twirled Readout Error Extinction (T-REx) and dynamical decoupling (DD)\cite{pokharel2018demonstration, van2022model}. Central to this refined approach is the utilization of a Noisy Intermediate Scale Quantum (NISQ) source device for heightened performance \cite{nisq}. Extensive tests on both simulated and real-world quantum devices, notably IBM’s quantum computer, provide compelling evidence of the effectiveness of our T-REx and DD adaptations in mitigating prevalent quantum channel errors.

Subsequent to the entanglement distribution, our protocol adopts a verification method reminiscent of Quantum Key Distribution (QKD) schemes. The Commander then issues orders encoded in specific quantum states, like ``Retreat" or ``Attack". In situations where received orders diverge, lieutenants engage in structured games to reconcile discrepancies. Notably, the frequency of these games is contingent upon the Commander's strategies and the overall network size. Our empirical findings underscore the enhanced resilience and effectiveness of the protocol in diverse scenarios. Nonetheless, scalability emerges as a concern with the growth of the network size. To sum up, our research illuminates the considerable potential of fortified quantum consensus systems in the NISQ era, highlighting the imperative for sustained research in bolstering quantum ecosystems.

\end{abstract}

\keywords{Suggested keywords}

\maketitle

\section{Introduction}
Distributed computing, a pivotal aspect of modern computational systems \cite{turek}, often grapples with the complexity of consensus building, especially when viewed through the lens of quantum technology \cite{buhrman2003distributed,caleffi2022distributed, van2016path}. A specialized variant of the Byzantine Agreement (BA) problem—Detectable Byzantine Agreement (DBA)—presents itself as a significant challenge yet an avenue for showcasing the potential of resources endemic to quantum computers in fostering consensus.\cite{gaertner2008experimental} Our exploration uniquely navigates scenarios where classical consensus building remains distinctly unresolvable, particularly when greater than one-third of processes either fail or act in adversarial manners. Within the realm of quantum distributed computing, we propose a novel protocol that can accommodate any number of faulty or adversarial players even amidst channel errors of typical current generation quantum computers. This proposition showcases resilient error mitigation (EM) and stability amidst adversarial coordination, heralding a significant stride towards robust quantum-based consensus systems.

The realm of Quantum Key Distribution (QKD) has been illustrated across numerous applications, marking a pivotal stride in secure communication\cite{scarani2009security}. Furthermore, in the emerging NISQ era of quantum technological development, various error correction and mitigation techniques are employed to manage the inherent fragility and noise in quantum systems\cite{russo2023testing, kim2023evidence,GHZQEM2023quantum}. Our exploration is funnelled into two distinct error suppression techniques from IBM's Qiskit Runtime Service: DD and T-REx\cite{pokharel2018demonstration, van2022model}. While DD injects additional pulses to idle qubits to attenuate decoherence effects, T-REx exploits Pauli twirling to mitigate noise during quantum measurement without necessitating a particular noise form, showcasing its generic applicability and effectiveness.

As we journey through the quantum technological landscape, it is imperative to contextualize our discussion within the burgeoning realm of distributed quantum computing and the quantum internet\cite{wehner2018quantum}. The latter, an innovative confluence of remote quantum devices, leverages both quantum and classical links, heralding capabilities unparalleled by classical analogues, such as quantum teleportation and entanglement \cite{kimble2008quantum,cuomo2020towards}. This enthralling advent, grounded firmly within the principles of quantum mechanics, brings forth novel network design constraints and enables computational feats beyond the reach of classical systems \cite{cacciapuoti2019quantum}. The coalescence of the Quantum Internet and distributed quantum computing holds the potential to exponentially enhance quantum computing power through scalable networks of qubits. \cite{bravyi2022future,caleffi2018quantum}

Amidst advancements in quantum computing arises a rising imperative for fortified decentralized trust management systems, especially within distributed cyber-systems where veracity is indeterminate. Traditional centralized trust management systems, entangled with the predicaments of a zero-trust environment, gradually give way to blockchain technologies. The latter meticulously crafts decentralized trust through consensus mechanisms, adeptly navigating challenges conjured by extensive data sharing and consensus \cite{qu2023quantum}. The impending threat quantum technology places upon classical blockchain systems evoke a shift towards a paradigm embodied by n-party quantum DBA, instantiated through the GHZ state, signifying a groundbreaking stride towards facilitating data consensus within a quantum blockchain \cite{yang2022decentralization}.  This fusion of quantum computing and blockchain technology heralds a new framework, amplifying computational power while securing decentralized networks in a zero-trust landscape, potentially revolutionizing data management and expanding opportunities for scientific and academic advancements.

\begin{figure}[htbp]
\centering
\includegraphics[width=0.45\textwidth]{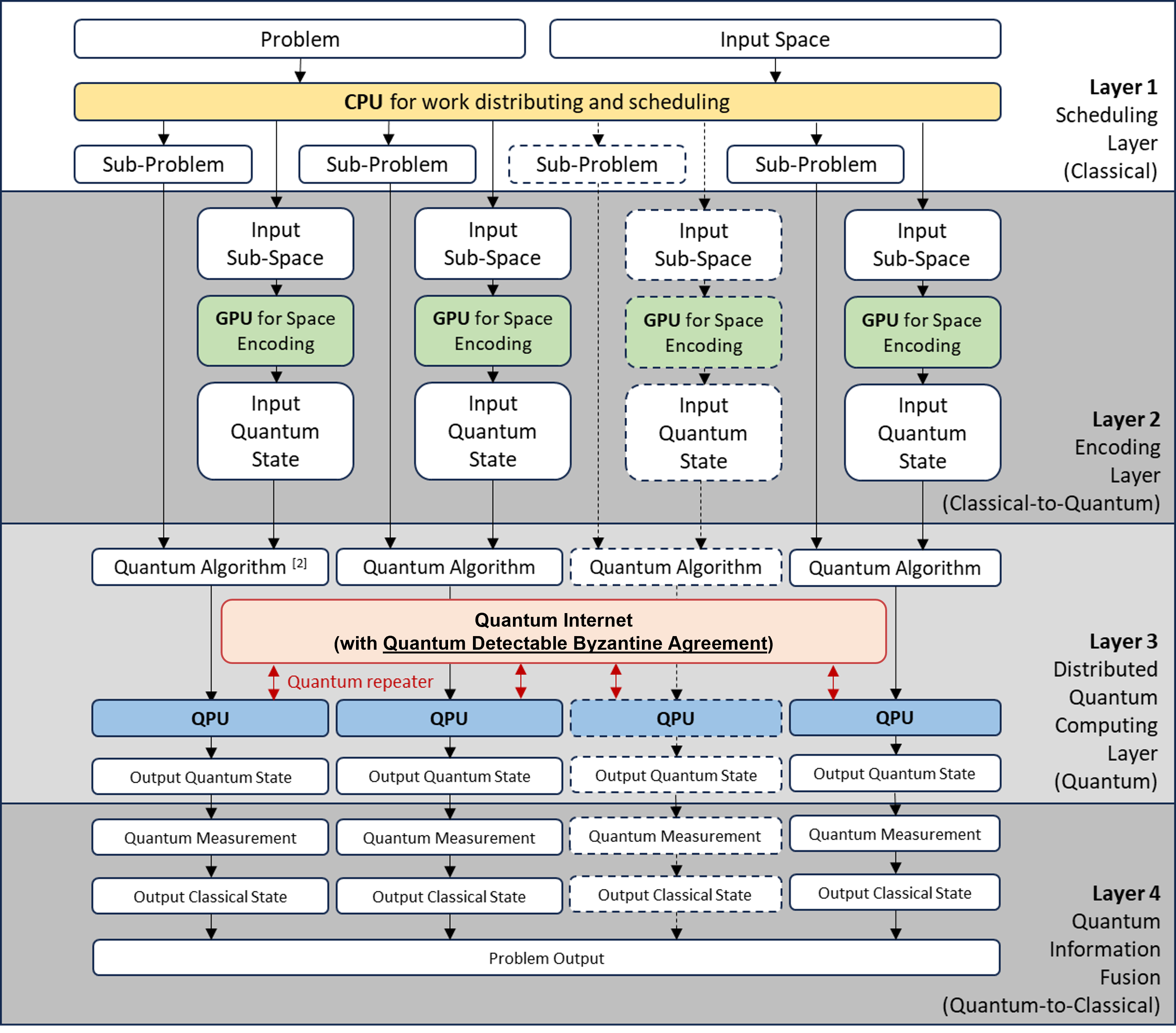}
\caption{Schematic representation of a distributed quantum computing architecture, illustrating the integration of classical scheduling with quantum processing through various layers, encompassing space encoding, quantum algorithms execution, quantum internet communication, and the transition from quantum to classical information.}
\label{fig1}
\end{figure}

 In Fig. \ref{fig1}, we outline a distributed quantum computing architecture which begins at the classical Scheduling Layer. A principal computational challenge is segmented into multiple sub-tasks by a dedicated central processing unit (CPU) to facilitate efficient task distribution and scheduling\cite{topcuoglu2002performance}. The ensuing sub-tasks proceed to the Encoding Layer, where graphic processing units (GPUs) convert classical input sub-spaces into quantum-adaptable formats, resulting in input quantum states\cite{bayraktar2023cuquantum}. These states advance to the Distributed Quantum Computing Layer, where they are acted upon by specialized quantum algorithms, facilitated by quantum processing units (QPUs)\cite{cuomo2020towards}. To ensure communication among QPUs, an advanced quantum internet framework, bolstered by the quantum DBA, is incorporated to ensure a robust and secure quantum data exchange \cite{van2022quantum, taherkhani2017resource}. Quantum repeaters are pivotal in maintaining quantum coherence during transmissions, culminating at the quantum information fusion Layer, where resultant quantum states undergo measurements, converting them to classical states \cite{taherkhani2017resource}. The aggregated classical outcomes from multiple QPUs are then amalgamated to derive the final solution to the primary computational challenge.

The architecture propounded here provides an innovative modus operandi for enhancing the scalability of qubit numbers in quantum systems. By distributing computational processes across diverse QPUs and leveraging a quantum internet framework, it enables parallel processing and the integration of an extensive qubit array, transcending the capabilities of conventional, centralized quantum systems\cite{bravyi2022future}. In the NISQ era, characterized by intermediate qubit counts prone to errors, this scalability is crucial. As the architecture fosters qubit proliferation, it concurrently enhances the efficacy and flexibility of quantum algorithms. This elevation, attributed to expanded quantum capacity, enables the exploration and execution of sophisticated quantum algorithms\cite{bharti2022noisy}, offering significant computational advantages, and easing the transition from the constraints of the NISQ era to the potential of full-scale, error-resistant quantum computing.

Furthermore, this distributed architecture engenders a conducive ecosystem for the implementation of Quantum Federated Learning (QFL), whereby a collective learning model is built from decentralized datasets residing on distinct quantum devices, enhancing the privacy-preservation and efficiency of machine learning processes across the network \cite{rofougaran2023federated, chehimi2022quantum}. It also facilitates advanced methods like Accelerated VQE for more efficient quantum chemistry computations, optimizing the trade-offs between runtime and circuit depth, especially relevant in near-term quantum machines\cite{diadamo2021distributed}. Large-scale Quantum Approximate Optimization Algorithm (QAOA) implementations, pivotal in solving combinatorial optimization problems, find fertile ground in this architecture, bolstering the expedition of solutions for real-world optimization challenges by distributing the computational load across multiple QPUs \cite{zhou2023qaoa}. IBM's quantum-centric supercomputing endeavours dovetail with this architecture, embodying a visionary stride towards integrating classical and quantum computing resources for high-performance computing applications \cite{bravyi2022future}. The synthesis of Quantum Internet and distributed quantum computing, as delineated in this architecture, paves the path for an orchestrated quantum-classical computational paradigm, augmenting the computational prowess and opening new vistas for quantum algorithm developments. The burgeoning collaborations between quantum computing and blockchain technology, underscored by this architecture, herald a promising frontier for secure, decentralized data management, and consensus systems in a quantum-enabled digital ecosystem.

\par

\section{Protocol Structure}

\begin{figure}[htbp]
\centering
\includegraphics[width=0.45\textwidth]{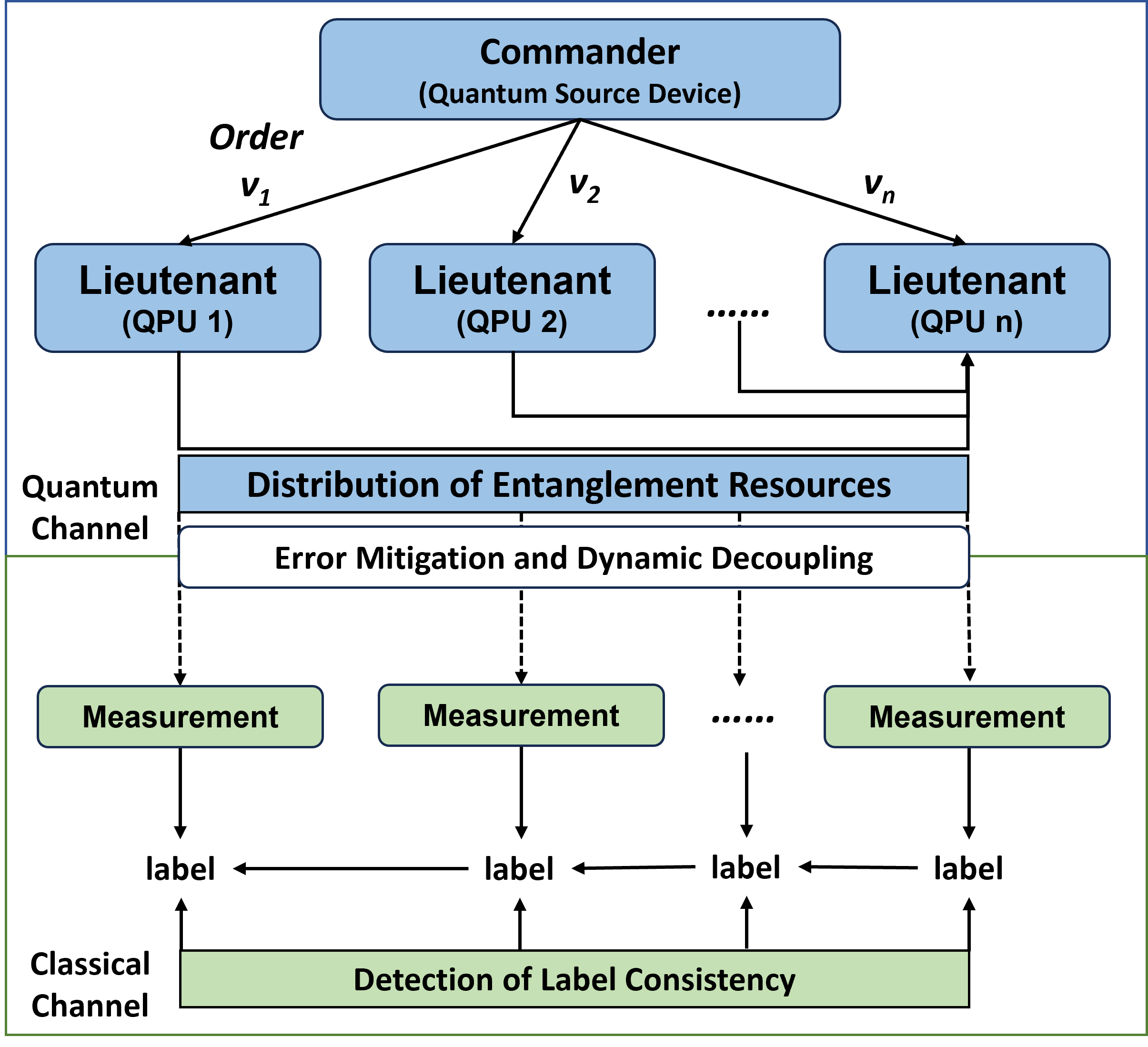}
\caption{Flowchart depicting the quantum DBA protocol with integrated error-mitigation and dynamic decoupling strategies. The structure emphasizes the roles of the Commander (quantum source device) and Lieutenants (quantum processing units), followed by processes for entanglement distribution, error correction, quantum measurements, and the final step of label consistency detection.}
\label{fig2}
\end{figure}

In the realm of a distributed quantum computing network, especially in scenarios dealing with QBA, it's crucial that QPUs spread across various nodes reach a consensus regarding the data to be recorded in the subsequent block. In Fig. \ref{fig2}, the proposed quantum DBA protocol delineates a method to achieve this, ensuring the integrity and reliability of the agreement process, even in the presence of potentially dishonest participants. Below is an in-depth explanation of the steps involved in this protocol:

\begin{enumerate}
    \item \textbf{Distribute quantum resources to all generals}:
    In the initial phase, quantum resources (which can include entangled qubits) are distributed among all participant nodes, referred to as ``generals" in the Byzantine Agreement context \cite{cholvi2022quantum}. This distribution might be achieved through quantum channels, ensuring secure communication and the establishment of a shared quantum state among participants. This shared state is critical as it forms the basis of the subsequent quantum communication and entanglement-assisted agreement processes.
    
    \item \textbf{Verify entanglement correlations:}
    Once the quantum resources are distributed, participants undertake the verification of entanglement correlations to ascertain the expected entanglement of the shared quantum states. This critical step may encompass procedures such as Bell tests or alternative quantum state tomography techniques (e.g., GHZ state\cite{qu2023quantum}), empowering the generals to affirm that the initial quantum resources retain a state conducive to secure and dependable communication. This verification functions as a protective measure against potential eavesdropping or tampering with the quantum channels.
    
    \item \textbf{Commander issues orders to lieutenants:}
    The commander utilizes the established quantum channel to issue orders to the other participants, known as lieutenants. These orders are transmitted securely through quantum states, and due to the properties of quantum entanglement, they cannot be intercepted or altered without detection. Each lieutenant receives the order and stores it for the agreement process.
    
    \item \textbf{Agreement/disagreement resolution between lieutenants:}
    Upon receiving orders, lieutenants engage in a vital consensus-building phase, leveraging their entangled states for secure discussions about the commander's directives. This phase may invoke protocol-allowed quantum operations to identify and isolate deceptive nodes.\cite{gaertner2008experimental} Lieutenants undertake thorough quantum communications, using advanced techniques like quantum voting\cite{xue2017simple}, superdense coding\cite{harrow2004superdense}, or entanglement-swapping\cite{zukowski1993event}, to mitigate deceit and verify order authenticity. If consensus is achieved, they proceed as agreed; otherwise, intensified discussions or conflict resolutions are invoked in deceitful scenarios. Persistently dishonest or non-cooperative nodes, identified through quantum methods, are excluded from future dialogues to maintain process integrity.
\end{enumerate}

By implementing these steps, the quantum DBA protocol utilizes quantum mechanics to ensure secure, efficient, and dependable consensus in distributed networks, resilient to threats from dishonest participants. This approach is particularly potent in situations where traditional methods may stumble due to security flaws. Importantly, the integration of QEM enhances the fidelity of states, crucial for maintaining operational integrity amidst quantum computing's susceptibility to errors. Concurrently, strategies to counteract DD are employed, preserving the coherence of quantum states, and thereby bolstering the reliability and accuracy of the consensus process. These advanced techniques collectively strengthen the protocol, making it a formidable solution in the evolving landscape of quantum communications.

\subsection{Distribute Entanglement}
Let $n$ be the total number of generals $(n > 2)$, including the Commander and all Lieutenants. Let $k$ be the number of copies of the quantum resource distributed amongst the $n$ generals. The entangled state distributed from the Quantum Source Device (QSD) is defined as
\begin{align}\label{PSI}
    \ket{\Psi_n} \equiv \frac{1}{\sqrt{3}}\bigg[\ket{00}\ket{1}^{\otimes n-1} + \ket{11}\ket{0}^{\otimes n-1} + \\ 
    \frac{1}{\sqrt{2(n-1)}} \bigg( \ket{01}\sum_{i=0}^{n-2}\ket{0}^{\otimes n-2-i}\ket{1}\ket{0}^{\otimes i} \nonumber \\ 
    + \ket{10}\sum_{i=0}^{n-2}\ket{1}^{\otimes n-2-i}\ket{0}\ket{1}^{\otimes i}\bigg) \bigg]. \nonumber
\end{align}

This state is distributed such that the commander receives the first two qubits and all lieutenants receive a single qubit. This state has the following measurement outcome correlations:
\begin{enumerate}
    \item If the commander measures $\ket{\psi_{com}} = \ket{00}$, then all lieutenants will measure $\ket{1}$.
    \item If the commander measures $\ket{\psi_{com}} = \ket{11}$, then all all lieutenants will measure $\ket{0}$.
    \item If the commander measures $\ket{\psi_{com}} = \ket{01}$, then one lieutenant at random will measure $\ket{1}$ while all others measure $\ket{0}$.
    \item If the commander measures $\ket{\psi_{com}} = \ket{10}$, then one lieutenant at random will measure $\ket{0}$ while all others measure $\ket{1}$.
\end{enumerate}
Note that state $(\ref{PSI})$ is symmetric under exchange among the lieutenants.

\subsection{Verify Entanglement}

\begin{figure}
    \includegraphics[width=\linewidth]{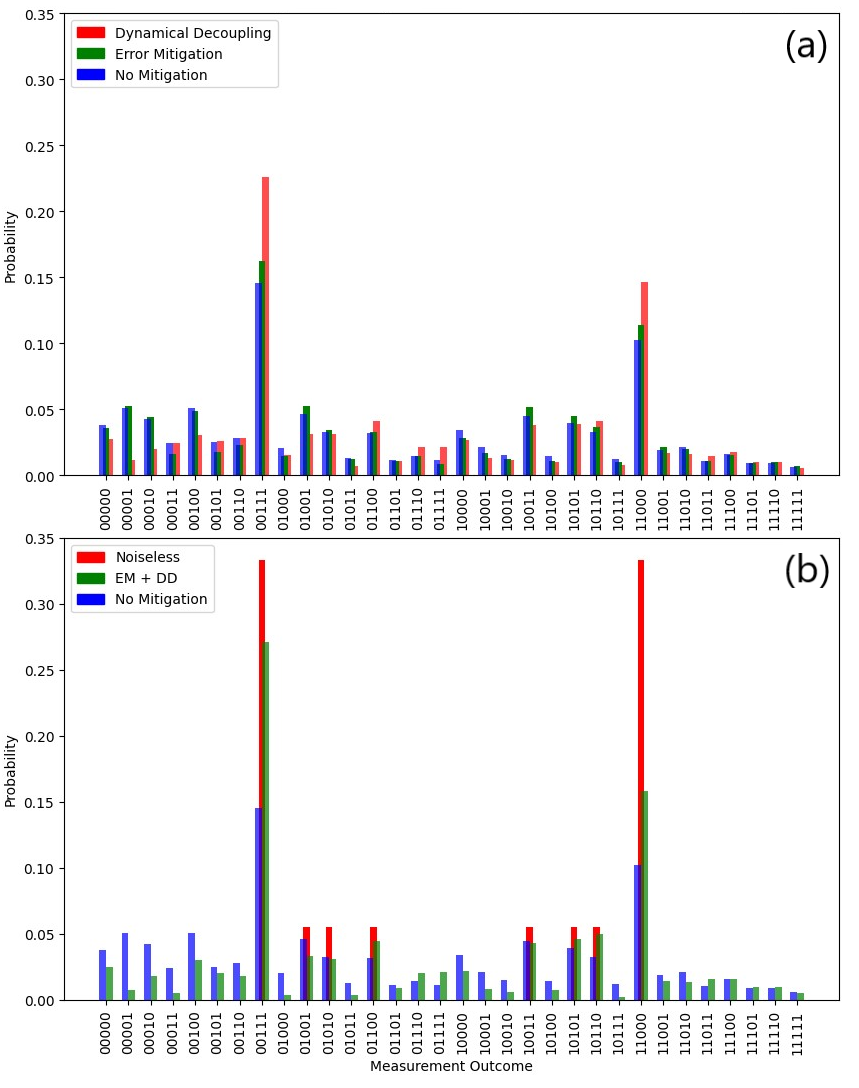}
    \caption{$(a)$ And $(b)$, respectively depict the measurement outcomes of the initial state with and without EM (T-REx) and DD. The depicted probability mass functions are the averages of $100$ iterations of measuring the prepared state $(\ref{PSI})$ with $n = 4$, on IBM's Nairobi quantum computer available via Qiskit Runtime with each iteration containing $8192$ shots.}
    \label{results0}
\end{figure}

Once $k$ copies of state (\ref{PSI}) have been distributed and indexed such that the commander receives the first two qubits of every copy and all other players receive a single remaining qubit of every copy the verification can begin. A subset of indices $i < k$ are selected and all generals exchanged values pairwise recording the correlations for later reference in the same manner as Quantum Key Distribution (QKD) schemes.

Fig. \ref{results0} illustrates the entanglement resource (\ref{PSI}) under various conditions. In Fig. \ref{results0}$(a)$ the impact of EM and DD is evident in the increased probability of measuring the outcomes corresponding to \textbf{retreat} and \textbf{attack} as compared to the no mitigation scenario. In Fig. \ref{results0}$(b)$ the combined effect of EM and DD is further evident in its proximity to the noiseless case. Interestingly the difference in probability of the \textbf{retreat} and \textbf{attack} outcomes in the no mitigation case is magnified when EM and DD are applied, the equality between these two outcomes that exists in the noiseless case is not recovered.

\subsection{Distribute Order}
The commander issues an order mapped by $\ket{00} = \text{Retreat, } \ket{11} = \text{Attack}$. The order along with a subset of remaining indices $j < k - i$ corresponding to the order are sent along pairwise classical channels to all lieutenants.

\subsection{Realize Agreement}
All lieutenants then exchange their received orders pairwise with one another along classical channels. Any disagreements initiate a pairwise turn-based game with the following structure:
\begin{enumerate}
    \item Random starting lieutenant sends index with corresponding order they received.
    \item The receiving lieutenant constructs a joint distribution of the opponents claimed order value and their own value at the corresponding received index.
    \item The receiving lieutenant then takes a turn sending their claimed order and an index corresponding to that order.
    \item After a number of indices $l$ are exchanged, the pairwise game is terminated.
    \item After all pairwise games between disagreeing lieutenants are complete, each lieutenant then compares their receiving joint distributions to the distributions obtained during the verification step, classifying them as either closer to the commander fail rate or to the traitor fail rate.
    \item If a loyal lieutenant who received an attack order detects that the commander is a traitor, they switch their action to retreat.
    \item All loyal lieutenants' actions are collected and the DBA criteria is applied.
    each player calculates the failure rate $F/l$ of their opponent. If $F/l$ is above a predetermine, noise-dependent threshold $T$, the opponent is declared a traitor and no action change occurs. If $F/l$ is below $T$, the commander is declared a traitor, the player then sets their own action to retreat.
\end{enumerate}

How many turn-based games occur as a function of the number of generals and the number of traitors? With the assumption that a traitorous lieutenant always claims to have received the opposite order we can determine the number of pairwise games that will occur. With this state structure for $n$ players and $m\leq n$ traitors, the number of pairwise games that occur if the commander is loyal becomes
\begin{align}
    g = m(n-1-m).
\end{align}
If the commander is a traitor, the number of pairwise games needed can be derived by maximizing $g$ with respect to $m$
\begin{align}
    g_{max} = \frac{(n-1)^2}{4}.
\end{align}
In essence, the number of games that occur when the commander is a traitor depends on how the commander selectively disseminates orders. This could induce any number in the interval $[0, g_{max}]$. A traitorous commander seeking to minimize the likelihood of achieving DBA by giving half of the generals opposite orders, achieving $g_{max}$ number of pairwise games. As $g$ scales as $n^2$, the number of games and therefore protocol computation time scales rapidly with additional players.

In the simplest case, a turn-based game between a traitor and a loyal general who both received an \textbf{attack} order from a loyal commander has the following structure:
\begin{enumerate}
    \item The loyal lieutenant has no available decisions to make, they simply send the indices received from the commander.
    \item The traitor rationally chooses to optimize their options to increase the probability of being undetected.
    \item the options available are the remaining indices corresponding in which the traitor has $\ket{0}$ or $\ket{1}$.
    \item To maximise the probability of deception the traitor will select from remaining indices in which they have the same value as the order in which they received. When doing so, the traitor failure rate in the absence of noise becomes
    \begin{align}
        R_{TD} = \frac{2}{3(n-1)}.
    \end{align}
\end{enumerate}

\subsubsection{Deriving the Traitor Failure Rate}
Without loss of generality the Traitor Detection Rate $R_{TD}$ is defined as
\begin{align}
    R_{TD} \equiv 1 - P(L = 1 | T = 1),
\end{align}
where $L$ is the outcome of measurement for a particular index of the loyal lieutenant and $T$ is the outcome for the traitor.
\begin{align}
    \implies R_{TD} = 1 - \frac{P(L = 1 \cap T = 1)}{P(T = 1)}.
\end{align}
As \ref{PSI} is identical under exchange of lieutenants, without loss of generality
\begin{align}
    P(T = 1) = P(\ket{q_2} = \ket{1}), \\
    P(L = 1) = P(\ket{q_2} = \ket{1}),
\end{align}
where $\ket{q_2}$ and $\ket{q_3}$ are the third and fourth qubits in the array respectively.
\begin{align}
    \implies P(T = 1) = \lvert\frac{1}{\sqrt{3}}\rvert^2 + \lvert\frac{1}{\sqrt{6(n-1)}}\rvert^2 + ...
\end{align}
\begin{align}
    = \frac{1}{3} + (n-1)\lvert\frac{1}{\sqrt{6(n-1)}}\rvert^2.
\end{align}
\begin{align}
    \implies P(T = 1) = \frac{1}{2}.
\end{align}
While,
\begin{align}
    P(L = 1 \cap T = 1) = \lvert\frac{1}{\sqrt{3}}\rvert^2 + \prod_{i=0}^{n-4}\lvert\frac{1}{\sqrt{6(n-1)}}\rvert^2.
\end{align}
\begin{align}
    \implies P(L = 1 \cap T = 1) = \frac{1}{2} - \frac{1}{3(n-1)}.
\end{align}
Thus,
\begin{align}
     R_{TD} = \frac{2}{3(n-1)}.
\end{align}

\begin{figure}[htbp]
    \includegraphics[width=\linewidth]{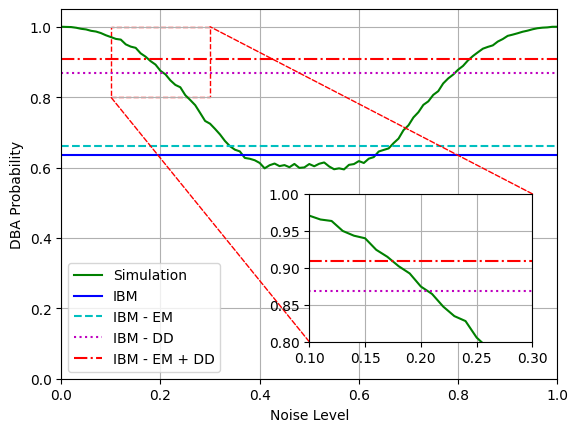}
    \caption{The probability of achieving detectable broadcast in the presence of noise, simulated as bit-flip error compared to using IBM Nairobi with number of indices exchanged (shots) equal to $1000$ in the case with the total number of generals $n = 5$ with a single, randomly selected traitor. The noise level represents the probability for any qubit value to flip upon measurement and ranges from $0$ to $1$ in $100$ increments with $10000$ iterations being performed at each value. As this scheme depends on the correlations between qubits, the probability of success is necessarily symmetric under reflection about the line where the noise level is $0.5$. This represents the point of no correlation, or maximum scrambling, as the noise level increases beyond this anti-correlation occurs which can also be used as an entanglement resource in an identical fashion. Under these parameters, the use of EM and DD is equivalent to a noise level of $0.175$, an improvement over the use of DD alone $(0.207)$, both of which are a dramatic improvement compared to the unmitigated results $(0.338)$.}
    \label{results1}
\end{figure}

\begin{figure}[htbp]
    \includegraphics[width=\linewidth]{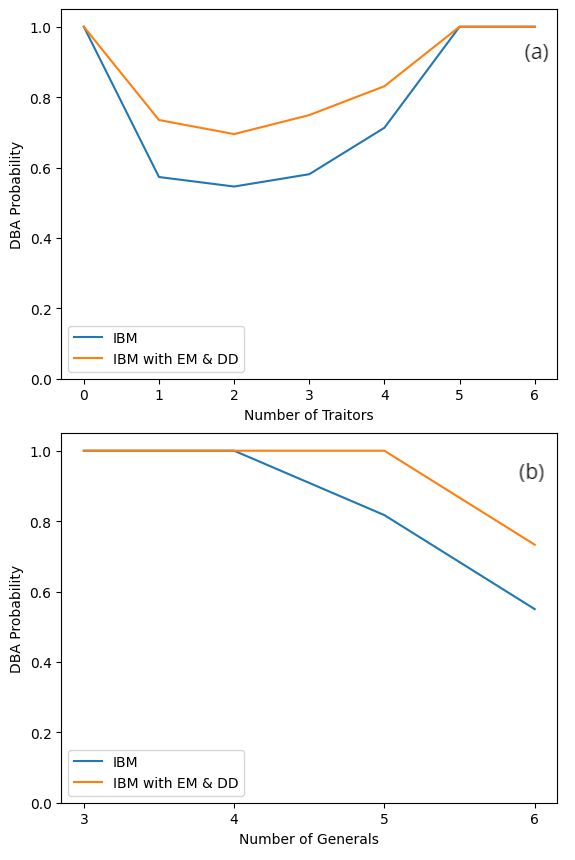}
    \caption{(a) Depicts the probability of achieving detectable broadcast with and without EM and DD with a fixed network size $n = 6$, and varying numbers of traitors present, while (b) depicts the probability of achieving detectable broadcast with and without EM and DD in the presence of a single traitor with varying network size. $1000$ iterations were performed for each data point, with a number of shots equal to $10,000$. The trivial cases in (a) with a number of traitors $m \in \{0,5,6\}$ are demonstrated clearly. In both (a) and (b) the significant performance improvement with the utilization of EM and DD is visible over the unmitigated case with greater success probability in every non-trivial case.}
    \label{results2}
\end{figure}

\begin{figure}[htbp]
    \includegraphics[width=\linewidth]{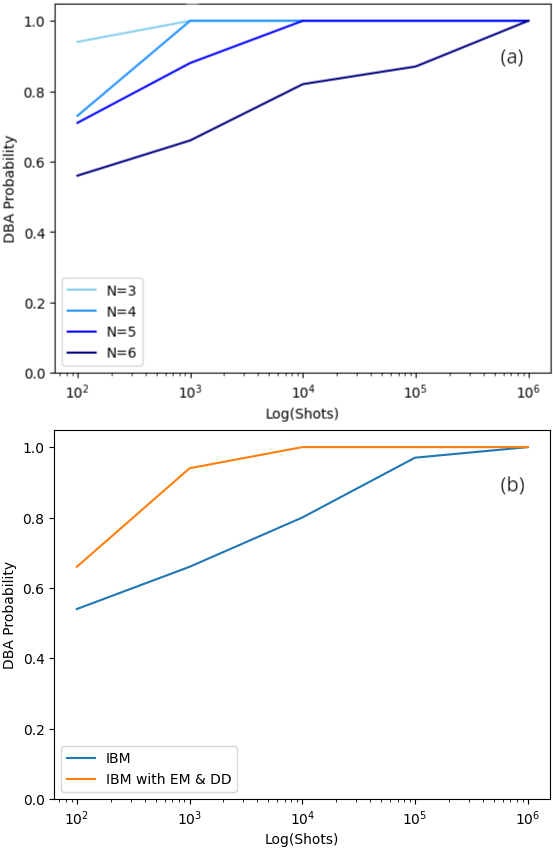}
    \caption{(a) Depicts the probability of achieving detectable broadcast with EM and DD for a range of different network sizes with increasing number of shots from $100$ to $10^6$. As the network size increases, the number of shots needed to achieve near-perfect broadcast, $P(DBA) \approx 1$, increases exponentially. Similarly, (b) depicts the probability of achieving detectable broadcast with and without EM and DD with a network size $n = 5$. The rate of convergence to near-perfect broadcast is significantly greater with EM and DD when compared to the unmitigated case. For both figures, $100$ iterations were performed at each data point.}
    \label{results3}
\end{figure}

\section{Results}
The proposed protocol was tested under varying conditions to determine the effective scaling of the scheme and assess the impacts of noise and the effectiveness of the error mitigation techniques examined. This exploration was limited to a maximum network size of $n = 6$ due to the maximum number of qubits available on chosen hardware.

In Fig.\ref{results1} as expected, the unmitigated case exhibits the lowest probability of success $(p_0 = 0.6368)$, while the addition of EM alone is a slight improvement $(p_1 = 0.6611)$. The presence of DD drastically improves the success of the scheme $(p_2 = 0.8689)$ even without the addition of EM. Interestingly, there is an apparent synergistic effect where the combined EM and DD success $(p_3 = 0.9093)$ shows a $1.663$ times greater impact from the EM than without DD. Even with a noise level of $0.5$ the probability of achieving detectable broadcast remains over $0.5$ for this scenario. This is expected as the traitor is chosen at random, as well as the order issued by the commander.

In Fig.\ref{results2} the impact of network size and number of traitors on the success of the protocol is explored. As expected, the addition of traitors reduces the probability of success up until approximately half of the lieutenants are traitors, from which point further addition of traitors reduces the number of loyal lieutenants needing to find consensus. The impact of network size also demonstrates the expected results of reduced success with increasing $n$. The number of shots used in Fig.\ref{results2} $(10,000)$ is sufficient to achieve perfect success $P(DBA) = 1$ up to $n = 5$ for the EM and DD case. The image of EM and DD is pronounced in adding robustness to this protocol for all cases explored here.

The scaling behaviour is explored in Fig.\ref{results3} with respect to the number of indices distributed (shots). This, along with network size affects the protocol computation time. The exponentially increasing number of shots required to achieve near-perfect success, with increasing network size, limits the scalability of this protocol as it operates currently. However, the impact of EM and DD does dramatically decrease the number of shots needed. Further research with larger systems and higher fidelity hardware will yield more conclusive evidence of the scalability of this protocol.

\section{Discussion}
In this paper, we delve into the intricacies of distributed quantum computing, emphasizing the challenges and potentialities of consensus building, especially when juxtaposed with quantum technology. A specialized variant, the DBA, emerges as both an obstacle and an opportunity, revealing the latent capabilities of quantum computers in consensus attainment. Our research uniquely explores realms where traditional consensus mechanisms falter, especially when more than a third of the processes are adversarial or faulty. We present a groundbreaking protocol for quantum distributed computing, capable of handling faulty or adversarial players amidst certain channel errors. This work also elucidates the interplay between QKD and quantum error correction techniques, emphasizing the efficacy of IBM’s Qiskit Runtime Service's DD and Twirled Readout Error Extinction. As the Quantum Internet materializes, merging with distributed quantum computing, it promises unprecedented capabilities, like quantum teleportation, transcending classical system limitations. The convergence of quantum computing with blockchain technology offers a formidable framework, amplifying computational prowess and securing decentralized systems. Our proposed architecture delineates a methodical approach for distributed quantum computing, enhancing scalability, and fostering quantum algorithm effectiveness. The proposed quantum DBA protocol, when tested, reveals the efficacy of error reduction techniques and the potential challenges in scalability. This study serves as a beacon, illuminating the path forward for quantum technology, distributed computing, and their intersection in the digital realm.

The next steps along this research endeavour are to test this protocol on larger hardware systems, using various other error suppression techniques to quantify and improve the scaling capacity of this protocol.

\section*{Acknowledgement}
The authors extend their gratitude to Roberto Bondesan for his invaluable discussions. This research was funded by the U.K. Engineering and Physical Sciences Research Council under Grant No. EP/W032643/1 and also received support from the IBM Quantum Researchers Program. K.C. is grateful for the financial support from both the Turing Scheme for the Imperial Global Fellows Fund and the Taiwanese Government Scholarship to Study Abroad (GSSA). Special acknowledgement goes to the Quantum Open Source Foundation and QuantumPedia AI for their contributions and their quantum computing mentorship program.

\nocite{*}

\bibliography{bibliography.bib}

\end{document}